\documentstyle[pra,aps,preprint,tighten,eqsecnum]{revtex}
\input{psfig}                                          
\begin{document}                                                                
\draft                                                                                   
\title{From Heisenberg matrix mechanics to EBK quantization: theory and 
first applications}

\author {William R.\ Greenberg, Abraham Klein, and Ivaylo Zlatev}
\address{Department of Physics, University of Pennsylvania, Philadelphia,
Pennsylavania, 19104-6396}

\author{Ching-Teh Li}  
\address{Department of Physics, National Taiwan University, Tapai, 
Taiwan 10764, R.\ O.\ C.}

\date{\today}

\maketitle  

\begin{abstract}
Despite the seminal connection between classical multiply-periodic motion
and Heisenberg matrix mechanics and the massive amount of work done on the
associated problem of semiclassical (EBK) 
quantization of bound states, we show, 
that there are, nevertheless, a number of 
previously unexploited aspects of this relationship that bear on the 
quantum-classical correspondence.  In particular, we emphasize a quantum
variational principle that implies the classical variational principle
for invariant tori. We also expose the more indirect connection between 
commutation relations  and quantization of action variables.  
In the special case of a one-dimensional
system a new and succinct algebraic derivation of the WKB quantization
rule for bound states is given. With the 
help of several standard models with one or two degrees of freedom, 
we then illustrate how the methods of 
Heisenberg matrix mechanics described  
in this paper may be used to obtain quantum solutions with a modest 
increase in effort compared to semiclassical calculations.  We also
describe and apply a method for obtaining leading quantum
corrections to EBK results.  Finally, we suggest several new or 
modified applications of EBK quantization.

\end{abstract}

\narrowtext   

\section{Introduction}                                                          

Though applications of great interest (and increasing complexity)
continue to be developed, e.\ g.,~\cite{1,2}, the theory of the semi-classical
quantization of invariant tori by the application of EBK quantization
conditions~\cite{3,4,5} appears to be a closed (or at least quiescent)
sector in the study of the relationship between the quantum mechanics and the
classical mechanics of non-separable systems.   
(For exceptions, see the work of Jaffe and collaborators ~\cite{5a,5b,5c}.
and the even more recent work of Morehead ~\cite{5d}.)
Focusing on systems with two degrees of freedom,
we can divide the many methods that have been developed and applied to this
subject into two main sub-categories, those based on the solution of 
Hamilton's equations as an initial-value problem, i.\ e., the calculation
of trajectories, and those based on algebraic methods involving trajectories
indirectly or not at all. Within this subdivision, referred to as M1 and
M2, respectively, we may distinguish principally: 
\begin{list}{M?}{\setlength{\labelwidth}{1.5in}
\setlength{\itemsep}{9pt} \addtolength{\leftmargin}{0.75 em}}
\item[M1a)]
Independent actions are computed from closed curves, signaling invariant
tori, generated by the intersection of trajectories 
on two independent surfaces of section~\cite{6,7,8}.
\item[M2a)]
If invariant tori exist, the dynamical variables can be represented 
as multiply-periodic functions of the angle variables, with Fourier coefficients
that depend only on the frequencies, or equivalently, actions. 
Equations of motion for the Fourier 
coefficients are obtained in either Hamiltonian or Lagrangian form,
and for each there is an associated variational principle.  The equations of 
motion are solved either perturbatively or non-perturbatively and the 
quantized actions and energy calculated in terms of the known
Fourier coefficients~\cite{9,10,11,12,13,14,14a,15,16}.  
This is the approach of paramount interest to us. 
\item[M1b)]
Consider the system Hamiltonian to be a sum of an integrable part and a 
perturbation.  The latter is turned on adiabatically over a time $T$.
From the assumption that the actions computed 
initially for the unperturbed system are approximate adiabatic invariants,
it is 
possible to obtain quantum energies~\cite{18a,18} (sometimes even for values
of the coupling strength at which associated invariant tori no longer 
exist).
\item[M2b)]
The Hamilton-Jacobi equation is solved iteratively, using a Fourier
series expansion, thus providing the generating function for the appropriate 
action-angle variables~\cite{17}.  (In contrast the Fourier series constructed
in M2a are to be understood as the explicit equations of transformation
from the original dynamical variables to the action-angle set.)
\item[M1c)]
It can be shown~\cite{19,19a} that a finite set of adjacent non-quantized
trajectories with the same total energy can be used to calculate accurate  
values of the actions for one of them.  
Quantized energies and associated actions are computed 
by linear extrapolation. In this method it is necessary to propagate 
trajectories until they almost close on themselves.
\item[M2c)]
Following the ideas of Birkhoff and Gustavson~\cite{50}, 
several groups~\cite{51,52,53,54} have carried out increasingly ambitious
programs for transforming a given Hamiltonian to normal form by a succession  
of canonical transformations.  The resulting Hamiltonian is 
quantized and energy values obtained.  
\item[M1d)]
Trajectories propagated over a sufficiently large multiple of the 
elementary periods of a multiply-periodic orbit can be Fourier 
transformed to yield the Fourier representation of the fundamental dynamical
variables.  Applying Percival's formulas~\cite{13}, 
the actions are computed.  Initial conditions are varied until 
quantized orbits are found~\cite{20,20a,20b}.    
\end{list}
For further discussion and a more exhaustive list of references, 
we refer the reader to Refs.~\cite{20,20a,20c}.      

Despite all this effort, there remains a gap in the 
study of the relationship between quantum mechanics and the theory of
invariant tori, especially as it relates to the work that utilizes the
description of invariant tori in terms of Fourier series.  
This assertion is grounded on the observation that each 
form of quantum mechanics is associated in a natural way with a corresponding
version of classical mechanics.  Thus in the same sense that Schr\"odinger
wave mechanics is naturally associated with the Hamilton-Jacobi equation, 
and the classical limit of the path-integral method is Hamilton's
variational principle,   
it is equally true 
that the classical limit of Heisenberg matrix mechanics for bound
systems are the equations of motion for multiply-periodic systems.  
Though some of the work cited is based, in the sense of the correspondence
principle, on this passage between quantum and classical descriptions,  
and the passage itself is in some respects well known (see ~\cite{5a,5b,5c}, 
for instance), the main thesis of this paper is that this route for passing
from the quantum to classical theory has not been fully explored.
A possible explanation is that although the basic correspondence is
well-known, it is hardly known outside the circle of the authors
and former associates that matrix mechanics can be derived
from a novel variational principle \cite{27,28,29,30}     
called the trace variational principle (see also \cite{30a})  
and that the classical limit of this principle is the variational
principle for invariant tori \cite{16}.  

Guided by this relationship, the main 
purpose of this work is to study the transition
from the quantum to the classical domain more thoroughly 
than heretofore for both the dynamics
and the kinematics (quantization conditions).  This has led to   
some results that we find it difficult to believe are
not known, but for which we do not have an independent reference.  
In Sec.\ II, we supply a brief but self-contained account of 
our version of matrix mechanics, with emphasis on the variational 
basis.  The passage to the semi-classical
limit is then carried out in Secs.\ III and IV, where we show that the
limit of the quantum variational principle is the variational principle
for invariant tori.  We find, however, that the relationship between the 
commutation relations and the EBK quantization conditions is more indirect,
the former corresponding in the classical limit to the Poisson bracket
relations and the latter to the Lagrange bracket relations \cite{101}.
In Sec.\ V, we study the commutation relations {\it per se}, the most tangible
result being a new derivation of the WKB quantization condition, applicable
to one-dimensional systems.  Section VI is devoted to some illustrative
numerical studies.  We describe in turn and then apply 
algorithms for carrying out 
the semiclassical calculations, the associated matrix quantum
calculations, and a method for calculating directly quantum corrections
to the semiclassical result.  The need for the latter as a separate
approach arises from the fact that the matrix quantum calculations
are designed to give better results than the semiclassical one for 
low-lying states but do not go over in any limit to the EBK calculation.
Finally, Sec.\ VII contains several suggestions for
new ways to use the semiclassical quantization scheme. 
In a final section, we make some proposals for further work.  
Preliminary accounts of the main new theoretical results of this paper
can be found in \cite{40a,40b}.

\section{Matrix mechanics}  
\subsection{Variational principles and equations of motion}   

We begin with a brief description of a variational principle,
the so-called trace variational principle, from which one can derive
Heisenberg's form of quantum mechanics.  Though a version of this principle
was suggested more than three decades ago~\cite{27}, and subsequently 
several publications
have been devoted to its exposition and further development~\cite{28,29,30},
it appears to be largely unknown by the community at large. Except for one 
brief allusion, \cite{16}, it has not been applied to the problem at hand.  

Most of the important elements are already present for a system with one 
degree of freedom, and we therefore 
focus attention on the Hamiltonian         
\begin{equation}                                                                
H = \frac{1}{2}p^2 + V(x),\label{eq:2.1}                                        
\end{equation}                                                                  
with equations of motion                                                        
\begin{equation}                                                                
[x,H]= ip,  \label{eq:2.2}                                                      
\end{equation}                                                                  
\begin{equation}                                                                
[ip,H]= dV/dx \equiv V',  \label{eq:2.3}                                        
\end{equation}                                                                  
derived by utilization of the commutation relation $(\hbar = 1)$                
\begin{equation}                                                                
[x,p]= i.  \label{eq:2.4}                                                       
\end{equation}                                                                  

In practice we are usually concerned with the matrix          
elements of (\ref{eq:2.2})-(\ref{eq:2.4}) in the representation 
in which $H$ is diagonal, with eigenvalues $E_n$. Namely, 
\begin{eqnarray}                                                                
(E_n - E_m)x_{mn} & = & ip_{mn},\nonumber\\                                      
(E_n - E_m)ip_{mn} & = & (V')_{mn},  \label{eq:2.5}                             
\end{eqnarray}                                                                  
and                                                                             
\begin{equation}                                                                
[x,p]_{nm} = i \delta_{nm},\label{eq:2.6}                                       
\end{equation}                                                                  
where $x_{nm}=\langle n|x|m\rangle$.  We shall feel free
to use both notations interchangeably. 

In early work~\cite{31} we have shown, for polynomial potentials, 
how the energy differences and the matrix elements                
$x_{mn},~p_{mn}$ can be obtained from Eqs.\ (\ref{eq:2.5}) and 
(\ref{eq:2.6}).  
The primacy of these elements can be seen by using the completeness 
relation for the evaluation of matrix
elements of a product.   The                
eigenvalues themselves can be found by the direct evaluation of the             
expectation values                                                              
\begin{equation}                                                                
E_n = H_{nn} = \langle n|H|n\rangle = 
\sum_{n'} \frac{1}{2}|p_{nn'}|^2 + \langle n|V(x)|n\rangle,         
\label{eq:2.7}                                                                  
\end{equation}                                                                  
where the application of completeness is illustrated in the kinetic energy term.
Ultimately,  we shall be concerned with extending
the previous algorithms to the multidimensional case.

Before proceeding to the discussion of a variational formulation, we add a
few remarks about the implementation of the above formalism. 
Equations (\ref{eq:2.5}) and (\ref{eq:2.6}) are, to start with, an infinite
set of sum rules that must be satisfied by the exact eigenstates of the 
Hamiltonian.  Starting at any point in the spectrum, sums spread without
cutoff as far as the configuration space will allow.  To obtain closure, we
must make two kinds of approximation.  The first is that the matrix 
elements are rapidly decreasing functions of $k=|n-m|$, so that all matrix
elements with $k>k_{max}$ can be set to zero.  The second is that the matrix
elements are, for sufficiently large $n$ and $m$, slowly varying functions
of $n+m$, so that for some values sufficiently far from the center of 
interest in a given calculation we can set 
$\langle n+r|x|m+r\rangle =\langle n|x|m\rangle$ for $|r|\ll \frac{1}{2}
(m+n)$.   With these assumptions
Eqs.\ (\ref{eq:2.5}) and (\ref{eq:2.6}) reduce to a finite set of equations, 
where in addition to the retained matrix elements of $x$ and $p$, the additional unknowns are a set of elementary energy differences from which all other
energy differences can be composed.  To determine this collection of 
variables, it suffices to 
utilize all the available equations of motion, but only the diagonal 
elements of the commutator.  This is consistent with the result, proved
in the next subsection, that the off-diagonal elements of the commutator
are a consequence of the equations of motion.  
                                                                                
A natural question to ask is whether Eqs.\ (\ref{eq:2.5}) 
can be derived from a variational principle?  
Here we wish to treat the matrix elements         
of $x$ and $p$ as variables in the variational statement $\delta E_n
=\delta H_{nn} =0$.  
There are, however, two obstacles to such an endeavor:  (i) The matrix 
elements are not all independent.  (ii) The same     
matrix elements appear in different energy functionals.  Thus $p_{nn'}$         
occurs both in $H_{nn}$ and in $H_{n'n'}$.  
For which is it to be a variational parameter?                                                          
A solution to the second problem posed is to form an average of the             
stationary functionals. It turns out that in order to derive the 
equations of motion as given above, it is necessary to choose the most
symmetrical possible average, namely, the trace.  
Thus we require                                         
\begin{equation}                                                                
\delta \sum_{n} H_{nn} = \delta {\rm Tr}H = 0.  \label{eq:2.8}                  
\end{equation}                                                                  
A solution to the first problem is to impose all the possible kinematical       
constraints, namely                                                             
\begin{equation}                                                                
\delta[x,p]_{nn'} = 0.  \label{eq:2.9}                                          
\end{equation}                                                                  
Multiplying (\ref{eq:2.9}) by a Lagrange multiplier matrix $(-i)\Lambda_{n'n}$  
($\Lambda$ is Hermitian), we add the result to (\ref{eq:2.8}) 
and are thus led to a master variational principle                              
\begin{eqnarray}                                                                
0&= & \delta F \nonumber \\
& \equiv &\delta {\rm Tr}\left\{H - i \Lambda[x,p]\right\} \nonumber \\   
& = & \delta {\rm Tr} \left\{H - ip[\Lambda,x]\right\}    \nonumber \\ 
& = & \delta {\rm Tr}\left\{H + ix[\Lambda,p]\right\} .  \label{eq:2.10} 
\end{eqnarray}                                                                  
The several forms are equivalent because of the assumed cyclic invariance       
of the trace. (This is certainly unobjectionable in practice where the trace    
is taken over a finite dimensional vector space.)                               
                                                                                
Carrying out the unconstrained variation of (\ref{eq:2.10}) with respect to the 
matrix elements $x_{n'n}$ and $p_{n'n}$, keeping $\Lambda$ fixed, and           
using the explicit form (\ref{eq:2.1}) of $H$  we obtain the equations          
\begin{equation}                                                                
p_{n'n} = - i[x,\Lambda]_{nn'},   \label{eq:2.11}                               
\end{equation}                                                                  
\begin{equation}                                                                
(V')_{nn'} = i[p,\Lambda]_{nn'}.  \label{eq:2.12}                               
\end{equation}                                                                  
Because of the invariance of the trace with respect to choice of basis,         
the representation $|n\rangle$ 
is, at this point, arbitrary.  The most convenient     
immediate choice is the one in which the Hermitian operator $\Lambda$ is        
diagonal.  By comparing with the known equations of motion, we then identify    
$\Lambda$ as the Hamiltonian.  

The derivation of the equations of motion does not exhaust the consequences
that can be drawn from the trace variational principle.  We shall now 
demonstrate from this principle that the 
vanishing of the off-diagonal matrix elements of the 
canonical commutator is a consequence of the equations of motion, leaving
only the diagonal elements as independent kinematical conditions.  This 
result is consistent with the ``empirical" finding above that a dynamical
scheme is, in fact, completely determined by adjoining this one-dimensional
(diagonal) array of kinematical constraints to the equations of motion.

To derive the off-diagonal elements of (\ref{eq:2.6}), we make use of the       
invariance of the trace under an infinitesimal change of basis.  In the new      
basis, the Hamiltonian will not be diagonal, in general, and thus we must       
allow for a change in the Lagrange multiplier matrix.  We calculate             
\begin{equation}
0 = \delta F 
= {\rm Tr} \{ -i (\delta H) [x,p]\} ,  \label{eq:2.13}
\end{equation}
since all other contributions vanish in consequence of the 
equations of motion.  If we express the infinitesimal change of basis in the     
standard form                                                                     
\begin{equation}
 \delta |n> = -i \epsilon \Theta |n>  , \label{eq:2.14}  \end{equation} 
where $\epsilon$ is infinitesimal and $\Theta$ is Hermitian, 
we recognize that in a     
variation about the energy diagonal representation                              
\begin{equation} 
 \delta <n|H|n> = i \epsilon <n|[\Theta,H]|n> = 0 .  \label{eq:2.15}                           \end{equation} 
On the other hand non-diagonal elements                                         
\begin{equation}  
 \delta<n|H|m> \equiv <n|\delta H|m> \equiv \delta H_{nm} \label{eq:2.16}
 \end{equation}
can be assigned arbitrary infinitesimal values consistent with Hermiticity.     
From this and (\ref{eq:2.13}) we 
conclude that the off-diagonal elements of    
$[x,p]$ vanish.    

One additional result of great importance is that a solution of the 
dynamical scheme proposed above guarantees that the Hamiltonian is 
diagonal, $H_{n,n'}=0$, $n\neq n'$.  This result is derived in the Appendix.
                                                                                
In place of Hamilton's equations (\ref{eq:2.5}) it is often more
convenient to consider Newton's equation                                        \begin{equation}                         
 (E_{n} - E_{m})^{2} x_{mn} = (V')_{mn}. \label{eq:2.16a}  
\end{equation}  
This equation may be derived from its own variational principle by substituting
the first of Eqs.\ (\ref{eq:2.5}) into the previous functional $F$.  The 
result is a new functional $G$ that can be written in the alternative
forms     \begin{eqnarray} 
G&\equiv & {\rm Tr} \{ H - H [x,[H,x]]\} \nonumber \\  
&=& {\rm Tr} \{ - \frac{1}{2} [x,H][H,x]+ V(x) \}. \label{eq:2.17} 
\end{eqnarray}
In the second form, we recognize that $G$ is the negative of the 
trace of the Lagrangian.  
To obtain Newton's (or Lagrange's) equations one varies $G$ with respect to
the matrix elements of $x$, keeping the matrix elements of the Hamiltonian
fixed.  From the structure of (\ref{eq:2.17}), this means keeping the 
{\em energy differences} fixed.  This formulation of the quantum theory
is completed by adjoining a form of the canonical commutator from which
the momentum operator has also been eliminated, namely  
\begin{eqnarray}  
 \delta_{nm} &=& [x,[H,x]]_{nm}  \nonumber \\  
& =& \sum_{l} (2E_{l} - E_{m} - E_{n}) x_{ml} x_{ln}  . \label{eq:2.18}
\end{eqnarray}   
It should also be mentioned that the vanishing of the off-diagonal 
matrix element of the commutator can equally well be proved by paraphrasing
for the functional $G$ the argument presented for the functional $F$.

Since our main interest in this paper is in autonomous systems with at least 
two degrees of freedom, we must now describe how the previous considerations
are modified by this generalization.
We therefore consider a system with $N$ coordinates ${\bf x}=(x_1,...,x_N)$  
and a Hamiltonian of the form     \begin{equation}
H= \sum_i \frac{1}{2}p_i^2 + V({\bf x}). \label{eq:2.19}  \end{equation}
The functional $F$ from which Hamilton's equations are derived takes the
form             \begin{equation}
F = {\rm Tr} \{ H -i\Lambda\sum_i [x_i ,p_i]\} , \label{eq:2.20}  \end{equation}
where $\Lambda$ is once again identified as the Hamiltonian.  Lagrange's
equations are derived from the functional $G$, where
\begin{equation}
G= {\rm Tr}\{-\frac{1}{2}\sum_i [x_i,H][H,x_i] + V({\bf x})\}. \label{eq:2.21}
\end{equation}

Before providing any further details, we must mention the problem of 
labeling of the eigenvalues and eigenstates of $H$. In our earlier work 
\cite{16},
we ``naturally" assumed that the labeling could be done by a choice of 
$N$ integers ${\bf n}=(n_1,...,n_N)$ of which 
we could keep track as we tuned one
or more coupling parameters, starting from values for which the problem
was integrable.  The same assumption was discussed rather
more thoroughly by Percival \cite{10,11,12} who 
emphasized that the validity of this
assumption is coterminous, in the semiclassical limit, with the existence
of invariant tori. 

With this understanding, it is not necessary to write the equations of 
motion a second time, but only to remember that there is now one equation for
each value of $i$ and in each of these equations to replace the integer $n$
by the corresponding vector ${\bf n}$.  There does remain one question
to be addressed that will be of some importance to us later.  This is the 
question of whether we can deduce from the variational principle the 
separate vanishing, for each value of $i$, of the off-diagonal 
elements of the commutators $[x_i,p_i]$.  This follows from the fact that
the rows and columns of the matrix $\Lambda$ are each labeled by an 
$N$-dimensional vector, provided the set of non-vanishing matrix elements of
the product $x_i p_i$ is disjoint from the corresponding set for any other
choice of coordinate index.   This will be true for any model that we study.    
We have been somewhat cavalier in the present discussion, but the questions
that we have slighted, in particular, why other elements of the 
algebra do not appear in the constrained variational principle, 
will be considered in more detail in the discussion that now follows.  

\subsection{Commutation relations and equations of motion}  

The canonical commutation relations
must be constants of the motion; this means that their
time derivatives should vanish \cite{101a}.  
We show this for the class
of Hamiltonians under study.  Consider first   \begin{eqnarray}
\frac{d}{dt}[p_i,p_j]&=& -[\frac{\partial V}{\partial x_i},p_j]
-[p_i,\frac{\partial V}{\partial x_j}] \nonumber \\
&=&0.        \label{eq:crp}        \end{eqnarray}  
This calculation shows that in the energy-diagonal representation (assuming
no degeneracy) the commutator of two different components of momenta has
no off-diagonal matrix elements. If we choose the momenta to be imaginary
Hermitian operators, it follows that the diagonal elements of the commutator
also vanish.  Thus it is clear that the commutators $[p_i,p_j]=0$ may be omitted from the dynamical scheme.  

Utilizing the previous result that the off-diagonal elements of the commutator
$[x_i,p_i]$ vanish, we calculate
\begin{equation}
0=\frac{d}{dt}[x_i,p_i]=[x_i,\frac{\partial V}{\partial x_i}]. \label{eq:crq}
\end{equation}
Since $V$ is to a large extent arbitrary, we may safely conclude that the 
coordinates all commute with one another.

Combining the previous two results, we next verify that
\begin{equation}
\frac{d}{dt}[x_i,p_j]=[p_i,p_j] +[x_i,\frac{\partial V}{\partial x_j}]=0.
\end{equation}
Finally we check compatibility by the calculation
\begin{equation}
\frac{d}{dt}[x_i,x_j]=[x_i,p_j] +[p_i,x_j] =0.
\end{equation}

We have thus shown that the commutation relations are compatible with the 
equations of motion for the class of Hamiltonians under consideration.
For the practical problem of constructing a calculus based on Heisenberg
matrix mechanics, the consequence of our deliberations is that at most 
only the elements diagonal in the energy representation 
of the commutators of a coordinate and the corresponding 
momentum can enter, if we use all the available equations of motion.

\section{Semiclassical limit: mathematical preliminaries}  

The purpose of this section and the one to follow 
is to show that the semiclassical theory
of invariant tori is the ``natural" limit of the quantum theory of
the previous section.  We shall first collect in the form of lemmas 
some of the mathematical
statements that we need.  In the following we shall
use the notation ${\rm Lim}\;\langle {\bf n}|O|{\bf n}\rangle$ to signify the 
leading term in the semiclassical limit of the designated matrix element.
Here $O$ is generally a product of elementary operators.  We consider 
first the one-dimensional case and define for a real Hermitian operator, $A$,
\begin{equation}
A_k(n)\equiv \langle n-\frac{1}{2}k|A|n+\frac{1}{2}k\rangle =A_{-k}(n).
\label{eq:3.1}  \end{equation}
Notice that this quantity is an analytic continuation of a nearby
physical matrix element and is our definition of the semiclassical limit of the
matrix element $\langle n|A|n+k\rangle $.  As shown in a previous work 
\cite{15},
this choice can be used to provide a completely algebraic basis for the 
standard WKB quantization rule for a vibrational degree of freedom. 

We note that with the definition (\ref{eq:3.1}), we have
for the first two terms of a Taylor expansion,  
\begin{equation}
\langle n|A|n\pm k\rangle \cong A_k(n) \pm \frac{1}{2}k\frac{\partial A_k(n)}
{\partial n}.  \label{eq:3.2}    \end{equation}
Next, with the array of amplitudes $A_k(n)$, for fixed $n$ and varying $k$
we associate a formal Fourier series
\begin{equation}
A(n,\theta) = \sum_{k=-\infty}^{\infty}A_k(n)\exp(ik\theta), 
\label{eq:3.3}  \end{equation}
though in practice we shall always deal with severely restricted sums.
Below, we shall then make extensive use of the average
\begin{equation}
\langle ABC...\rangle \equiv (2\pi)^{-1}\int\,d\theta A(n,\theta)
B(n,\theta)C(n,\theta)...\;\;,      \label{eq:3.4}   \end{equation}
which is just the constant term in the Fourier series of the product.
    
With the above preliminaries, we are prepared to state and prove a
series of elementary propositions. \\
{\em Lemma 1}:
\begin{equation}
{\rm Lim}\;\langle n|AB|n\rangle =\langle AB\rangle. \label{eq:3.5}
\end{equation}
For the proof we write      \begin{eqnarray}
{\rm Lim}\;\langle n|AB|n\rangle&=&{\rm Lim}\sum_{k>0}^{k_{max}}
[\langle n|A|n+k\rangle\langle n+k|B|n\rangle \nonumber \\
&& +\langle n|A|n-k\rangle\langle n-k|B|n\rangle]. \label{eq:3.6}
\end{eqnarray}
Here we have assumed that $n$ is sufficiently large that the sum over $k$
can be extended far enough in both directions, to $k_{max}$, to obtain
numerical convergence.  This establishes the limited and non-rigorous sense
in which the word proof is to be understood both here and below.  
%What we have done is tantamount to an interchange of an order of limits,
%from the order lim $\hbar\rightarrow 0, t\rightarrow\infty$, the appropriate
%order for going to the small $\hbar$ limit in a quantum mechanical bound 
%  state problem, to the reverse order needed to put one at the classical limit.

If we now apply (\ref{eq:3.2}) in sequence
first to the matrix elements of $A$, for example, and subsequently to the 
matrix elements of $B$, we find
\begin{eqnarray}  
{\rm Lim}\;\langle n|AB|n\rangle& =&\sum_{{\rm all}\,k}  
A_k(n)B_{-k}(n)[1+O(n^{-2})]  \nonumber \\
&=&\langle AB\rangle[1+O(n^{-2})].   \label{eq:3.7}      \end{eqnarray}  
The error estimate arises from the assumption that a derivative with
respect to $n$ is of relative order $(1/n)$.  That the error is of second
order is a consequence of our choice of definition (\ref{eq:3.1}).
The sum on the right hand side of (\ref{eq:3.7}) has the value required 
by the lemma.

{\em Lemma 1a}:  With the same assumptions as before, Lemma 1 can be extended 
to a product of more than two factors,   
\widetext
\begin{equation}  
{\rm Lim}\;\langle n|ABC...|n\rangle = \langle ABC...\rangle 
=\sum \delta_{k_1 +k_2 +k_3 +...\;,0}A_{k_1}(n)B_{k_2}(n)C_{k_3}(n)... \;\;.  
\label{eq:3.8}    \end{equation}
\narrowtext  
The proof is carried out by ordering the various upward-going, downward-going
and mixed contributions to the multiple sum so that Eq.\ (\ref{eq:3.2}) can be
applied.

{\em Lemma 1b}:  The previous lemmas can be extended to the multi-dimensional 
case.  Extending the boldface notation now to designate, in addition to the 
quantum numbers ${\bf n}$, also the integer vector ${\bf k}$ for the   
components of a multiple Fourier series and the vector $\bbox{\theta}$   
for an array of angle variables,  we introduce a formal multiple Fourier
series,       \begin{equation}
A({\bf n},\bbox{\theta}) = \sum_{{\bf k}}A_{{\bf k}}({\bf n})
\exp(i{\bf k}\cdot\bbox{\theta}).    \label{eq:3.9}   \end{equation}
The lemma then applies to the average
\begin{eqnarray}
\langle AB\rangle &\equiv& (2\pi)^{-N} \int d\bbox{\theta} A({\bf n},
\bbox{\theta}) B({\bf n},\bbox{\theta})  \nonumber \\
& =& \sum_{{\rm all}\,{\bf k}}A_{{\bf k}}({\bf n})B_{{\bf -k}}({\bf n})
\label{eq:3.10} \end{eqnarray}   
and to corresponding multiple products.

{\em Lemma 2}:      \begin{eqnarray}  
{\rm Lim} \langle n|O|n+k\rangle &=& O_k(n) \nonumber \\  
&=& (2\pi)^{-1}\int d\theta\exp(-ik\theta) O(n,\theta).   \label{eq:3.11}
\end{eqnarray}   
Here $O$ is a product of two or more elementary operators, since for a 
single operator, the previous statement is only a combination of the 
definitions (\ref{eq:3.1}) and (\ref{eq:3.3}).  The same equation applies in 
boldface notation.  We shall not actually need this lemma, since its
application would be to obtaining the semiclassical limit of the equations
of motion directly.  Our procedure, however, will be to obtain that limit
for the variational principle and then derive the equations of motion from the 
latter.   

We do require expressions for limits of commutators.  We consider the usual
Poisson bracket,     \begin{equation}
[A,B]_{PB} \equiv \sum_i [ \frac{\partial A({\bf n},\bbox{\theta})}
{\partial \theta_i}\frac{\partial B({\bf n},\bbox{\theta})}{\partial n_i}
-\frac{\partial A({\bf n},\bbox{\theta})}{\partial n_i}\frac{\partial 
B({\bf n},\bbox{\theta})}{\partial \theta_i}].   
\label{eq:3.12}     \end{equation} 
We can now state \\
{\em Lemma 3}:
\begin{equation}
{\rm Lim} \langle {\bf n}|[A,B]|{\bf n + k}\rangle = 
i\langle [A,B]_{PB}\rangle_{{\bf k}},  \label{eq:3.13}   \end{equation}
where the notion on the right hand is understood as the ${\bf k}{\em th}$
Fourier component of the Poisson bracket (\ref{eq:3.12}).   
This lemma can be extended to multiple commutators.  In practice,
we shall only require the double commutator, for which we find\\  
{\em Lemma 3a}:
\begin{equation}
{\rm Lim} \langle{\bf n}|[A,[B,C]]|{\bf n+ k}\rangle =
(i)^2 \langle [A,[B,C]_{PB}]_{PB}\rangle_{{\bf k}}. \label{eq:3.14}
\end{equation}

We shall utilize these results to develop the dynamics and kinematics
of a semiclassical quantization scheme.  Before turning to this task,
we note an important consequence of the results in this section.  Whereas
in quantities such as $\langle {\bf n}|ABC|{\bf n}\rangle$, different base 
values of ${\bf n}$ will occur, it is a consequence of the elementary
reasoning applied above that in the expression Lim$\langle {\bf n}
|ABC|{\bf n}\rangle$ only the reference value of ${\bf n}$ occurs.  Thus in the  semiclassical limit a trace decomposes into a sum of independent terms.

\section{Semiclassical limit: dynamical and kinematical scheme}    
\subsection{EBK Scheme}

We now apply the reasoning and results of the previous section to obtain the
semiclassical limit of the (Lagrangian) form of the trace variational principle.
We are interested in a value of the vector ${\bf n}$ for which the 
corresponding energy and eigenstate may be described accurately by the 
semiclassical approximation.  We should then form the trace over a space of 
states extending in both directions in choice of ${\bf n}$ compared to the 
reference state.  From the results of the previous section, however, it follows
that for the purpose of deriving equations of motion 
we may suppress the trace, because in the limit considered, 
as we have already pointed out and in contrast to
the quantum case, the states decouple.  It thus suffices to focus 
attention on a given state.  (At the end of this section, however, we shall 
have occasion to restore the trace.)   

We consider first the kinetic energy $T$.  If $E({\bf n})$ is the energy of the
state identified by ${\bf n}$, then here we shall encounter the correspondence  
principle in the form     \begin{eqnarray}  
\omega_i({\bf n})&\equiv& E(n_1 ,...n_i +\frac{1}{2},...) - E(n_1 ,...n_i -
\frac{1}{2},...)    \nonumber \\  
&\cong & (\partial E({\bf n})/\partial n_i),
\label{eq:4.1}     \end{eqnarray}  
i.\ e., the classical frequencies ${\bbox \omega}({\bf n})
=(\omega_1 ,...,\omega_N)$
approach the quantum energy differences. We then find by a perfectly  
straightforward examination that 

\widetext  
\begin{eqnarray}
{\rm Lim}\, 2\langle {\bf n}|T|{\bf n}\rangle&=&{\rm Lim}\sum_i \langle {\bf n}|
[x_i,H][H,x_i]|{\bf n}\rangle  \nonumber \\   
&=& {\rm Lim} \sum_{i,{\bf k}}[E({\bf n+k})-E({\bf n})]^2 \langle {\bf n}|
x_i |{\bf n+k}\rangle\langle {\bf n+k}|x_i|{\bf n}\rangle  \nonumber \\
&=& \sum_{i,{\bf k}}({\bf k}\cdot {\bbox \omega})^2 x_{i,{\bf k}}({\bf n})  
x_{i,-{\bf k}}({\bf n}) 
=  \langle 2T\rangle .    
\label{eq:4.2}   \end{eqnarray}  
\narrowtext  
We note that the same result may be found by applying Lemma 3 for the classical
limit of a Fourier component of a commutator, provided we recognize 
that the classical Hamiltonian is a function only of the action variables
${\bf n}$. This follows from the fact that the formal Fourier series  
introduced to represent the solution of the classical problem defines a 
canonical transformation to the correct action angle variables (see below).

For the quantity just evaluated, as well as for the potential energy,
\begin{equation}
{\rm Lim}\, \langle {\bf n}|V|{\bf n}\rangle =\langle V\rangle, 
\label{eq:4.3}   \end{equation}
the relative error is of order $(1/n^2)$,  provided we use the definition
(\ref{eq:3.1}).    

If we combine Eqs.\ (\ref{eq:4.2}) and (\ref{eq:4.3}) we have
\begin{equation}
{\rm Lim}\, \langle {\bf n}|L|{\bf n}\rangle =\langle L\rangle. 
\label{eq:4.4}  \end{equation}
Since this expression is the semiclassical limit of a quantity stationary with
respect to variations of matrix elements of the coordinates, keeping energy 
differences fixed, we expect (\ref{eq:4.4}) to be stationary with respect to
variations of the associated Fourier components, keeping the frequency 
components fixed.  Conversely, requiring the stationary property,
\begin{equation}
\frac{\delta \langle L\rangle}{\delta x_{i,{\bf -k}}}   =0,
\label{eq:4.5}  \end{equation}
keeping the frequencies fixed, yields Lagrange's equations in Fourier component
form, namely,          \begin{equation}  
({\bf k}\cdot{\bbox \omega})^2 x_{i,{\bf k}}({\bf n})=
\frac{\partial\langle V\rangle}{\partial x_{i,_{\bf -k}}}.     
\label{eq:4.6}    \end{equation}

Considered from the purely classical point of view, the solution of Eqs.\
(\ref{eq:4.6}) may be undertaken from several standpoints.  The choice of real
Fourier coefficients, adhered to throughout the present development,
has already determined half of the initial conditions, namely, all components of the velocity vanish initially.   
If we then specify $N$ fundamental Fourier components of the
coordinates,  
we thereby define a scheme, provided a solution exists, 
that determines the remaining Fourier components 
as well as the $N$ frequencies.  An alternative scheme, more
integral to the variational principle, is to specify $N$ (incommensurable)
frequencies and calculate the Fourier amplitudes from the equations of motion.
In this paper, however, our interest will be in adjoining $N$ quantization
conditions to the equations of motion.  The resulting set, if it has a 
solution, must then determine the Fourier amplitudes and the frequencies.  

We apply the EBK quantization condition (restoring $\hbar$ for the instant)
\begin{equation} 
I_i =(n_i+\frac{1}{4}\alpha_i)\hbar,   \label{eq:4.7}   \end{equation}
where $\alpha_i$, the Maslov index, which has the value two for an uncoupled
vibrational degree of freedom and  
for the models studied in Sec.\ VII.  
The action $I_i$ is given by 
the equivalent expressions \cite{9,13,16},       \begin{eqnarray}   
I_i &\equiv & \frac{\delta \langle L\rangle}{\delta \omega_i} \nonumber \\
&=& \sum_{{\bf k}}k_i({\bf k}\cdot{\bbox \omega}){\bf x_k x_{-k}} \nonumber \\
&=& \langle {\bf p\cdot}\frac{\partial}{\partial\theta_i}{\bf x}\rangle\, ,  
\label{eq:4.8}        \end{eqnarray}    
where, in the last form, ${\bf p}$ is the momentum vector of the system.
This completes the formal statement of the semiclassical quantization  for
invariant tori.    

\subsection{Connection between EBK quantization and commutation relations}

We have, in fact, given two quantization schemes in this paper,  
the completely quantum one
consisting of Heisenberg's equations of motion adjoined to diagonal elements 
of the canonical commutation relations and the semiclassical one consisting
of the classical equations of motion in Fourier component form adjoined to
EBK quantization conditions.  But a part of the logical story is clearly
missing.  On the one hand, we have derived the classical equations of 
motion from the quantum ones using correspondence principle arguments
modified only minimally from historical forms.  On the other hand, although
EBK quantization can also be derived, up to the value of the Maslov index,
from the correspondence principle, what has not been made clear is the
relationship between the canonical commutation relations and the EBK 
quantization conditions.  Though it is hard to believe that this question
has not been addressed at some point in the history of quantum mechanics,
we have never encountered such a discussion in the literature.  
Therefore, for the sake of completeness, we include it at this juncture.  

Consider first the diagonal matrix element of the commutation relation     
\begin{eqnarray}
{\rm Lim}\,\langle{\bf n}|[x_i,[H,x_i]]|{\bf n}\rangle
&=&-i{\rm Lim}\,\langle{\bf n}|[x_i,p_i]|{\bf n}\rangle \nonumber \\
&=& \langle\sum_{j}[\frac{\partial x_i}{\partial\theta_j}
\frac{\partial p_i}{\partial n_j} -\frac{\partial x_i}{\partial n_j}
\frac{\partial p_i}{\partial\theta_j}]\rangle \nonumber \\
&=& \sum_{j,{\bf k}}\frac{\partial}{\partial n_j}k_j({\bf k}\cdot
{\bbox\omega})  
x_{i,{\bf k}} x_{i,-{\bf k}}.     \label{eq:4.9}     \end{eqnarray}
The basic result exhibited here, that follows from Lemma 3, is that the limit
of the diagonal element of the commutator is the time average of 
the corresponding Poisson bracket (PB).
On the other hand, we observe that the time average 
of the fundamental Lagrange bracket $\{\theta_i,n_i\}$ has the value
\begin{eqnarray}
\langle\{\theta_i,n_i\}\rangle& \equiv& \langle\sum_j[\frac{\partial x_j}
{\partial\theta_i}\frac{\partial p_j}{\partial n_i} - \frac{\partial x_j}  
{\partial n_i}\frac{\partial p_j}{\partial\theta_i}] \rangle \nonumber \\
&=& \frac{\partial}{\partial n_i}\sum_{j,{\bf k}}k_i({\bf k}\cdot{\bbox\omega})
x_{j,{\bf k}}x_{j,-{\bf k}}   \nonumber \\
&=& \frac{\partial}{\partial n_i}I_i =1.    \label{eq:4.10}   \end{eqnarray}  
Thus the EBK quantization condition is related to the Lagrange bracket 
(LB) rather than to the PB.

It is, of course, well-known \cite{101} that one set of 
fundamental brackets implies the other.  But for this purpose we must consider
the full brackets rather than just their time averages over the torus.
We illustrate the argument for one degree of freedom.  We thus wish to
show that the formal Fourier series     \begin{eqnarray}
x(n,\theta)&=& \sum_q x_q(n)\exp(iq\theta),   \nonumber \\
p(n,\theta)&=& \sum_q iq\omega x_q(n)\exp(iq\theta),  \label{eq:4.11} 
\end{eqnarray}  
satisfy the PB condition  \begin{equation}
[x,p]_{PB}=[x,p]_{\theta,n} =1.   \label{eq:4.12}   \end{equation}
provided, of course, that Eqs.\ (\ref{eq:4.11}) satisfy the equations of 
motion.  In that case the Fourier series represent the canonical transformation
between the given canonical coordinates $x,p$ and the exact angle-action
variables $\theta,n$.  By working out the Fourier series for the PB,   
requiring that the constant term be unity and that the other Fourier components
vanish, we find, with the help of a linear transformation of the summation
indices, that 
\widetext
\begin{equation}   
\delta_{q,0} =\sum_k[(k^2 -\frac{1}{4}q^2)x_{k+\frac{1}{2}q}\frac{\partial}
{\partial n}(\omega x_{-k+\frac{1}{2}q}) + (k-\frac{1}{2}q)^2\omega x_{-k
+\frac{1}{2}q}\frac{\partial}{\partial n}x_{k+\frac{1}{2}q}.
\label{eq:4.13}   \end{equation}
\narrowtext  
In the sum, $k$ takes on both integral and half integral values.
For $q=0$ this expression reduces to the one-dimensional form of (\ref{eq:4.9}).
The vanishing of the remaining Fourier components must be a consequence,
we suspect, of the vanishing of the off diagonal elements of the 
commutator, i.\ e.\ , it must be the classical limit of this property.    
This can be verified directly, using Lemma 3a.  As we emphasized at the 
beginning of this discussion, the equations of motion must be involved,
as indeed they are, in the vanishing of these off-diagonal elements.

Using the results derived in Sec.\ IIB, we can extend the detailed 
considerations just given to the multi-dimensional problem.  This is in fact
the interesting case, since in the one-dimensional case the PB and LB are
indistinguishable, a fact that will be used later  
to derive the WKB quantization condition from the commutation relation.  
The correct value of the Maslov indices must somehow also be implied in the
multi-dimensional derivation of the EBK quantization conditions.  The only
derivation we can supply at the moment is one based on starting with a
separable system and assuming adiabatic invariance as one turns up the 
coupling.  

It may be of some interest to derive the $q\neq 0$ part of 
(\ref{eq:4.13}) directly from the classical variational principle.  
For this purpose, we find it necessary to retain the full
structure of the quantum variational principle, so that in the classical
limit we have not only an average over a given torus, but in addition,
from the trace operation,  
an integral over all tori.  In other words,
we consider an average over phase space and indicate it by a double
bracket notation, e.\ g.,    \begin{equation}
\langle\langle L\rangle\rangle \equiv \int dnd\theta L(\theta,n),  
\label{eq:4.14}   \end{equation}
where the phase average of $L$ has the form, \begin{equation}  
\langle L\rangle = \frac{1}{2}i^2 \langle  (\frac{\partial x}   
{\partial\theta})^2 (\frac{\partial H}{\partial n})^2 \rangle -\langle V\rangle.
\label{eq:4.15}  \end{equation}
We now subject the system to an arbitrary infinitesimal canonical
transformation about the exact solution.  In consequence of the equations of 
motion, the only non-vanishing contribution comes from the first term of 
(\ref{eq:4.15}) and has the form  \begin{eqnarray}  
\delta \langle\langle L\rangle\rangle
&=&-\langle\langle (\frac{\partial x}{\partial\theta})^2\frac{\partial H}
{\partial n}\delta \frac{\partial H}{\partial n}\rangle\rangle \nonumber \\ 
&=&\langle\langle \delta H[\frac{\partial^2 H}{\partial n^2}(\frac{\partial x}
{\partial\theta})^2 +\frac{\partial H}{\partial n}
\frac{\partial}{\partial n}(\frac{\partial x}{\partial
\theta})^2]\rangle\rangle.   \label{eq:4.16}  \end{eqnarray}
Here the second line has been obtained by an integration by parts with
respect to $n$ with boundary terms dropped.  This requirement is 
clearly the classical analogue of the invariance of the trace, 
as is the requirement that (\ref{eq:4.16}) vanish.  Finally if we 
assume that $\delta H$ is expanded in a Fourier series 
with no constant term and otherwise arbitrary Fourier coefficients,
we are forced to the conclusion that all Fourier components of the expression
in square brackets, other than the constant term, vanish.  
It is in fact easy to see that this conclusion coincides with the 
corresponding statement contained in (\ref{eq:4.13}).  

It is, finally, important to refer to the classical limit of the property
that the quantum theory is formulated in the representation in which
$H$ is diagonal.  The limit of this property is precisely the condition
that when the canonical transformation to the exact action-angle variables
has been carried out, it is signaled by the vanishing of the Fourier series
for the classical $H$, other than the constant term. 
In applications, we shall utilize this 
requirement in both its quantum and classical aspects as a test of the 
convergence of our solutions.  For example, in the classical limit, the 
difference
\begin{equation}
\langle H^2\rangle - \langle H\rangle^2 = \sum_{k\neq 0}H_k H_{-k}
\end{equation}
should vanish.

We summarize in Table I the analogy that has now been fully established 
between Heisenberg matrix mechanics in the energy-diagonal representation 
and the theory
of invariant tori.  For the sake of simplicity, the notation appropriate 
to one degree of freedom has been utilized.  All entries have been described in
the preceding text.  Note that we have included the Hamiltonian rather than 
the Lagrangian form of the variational principle.

\begin{table}
\caption{The concepts of the theory of invariant tori as the classical limit
of the concepts associated with Heisenberg matrix mechanics in the 
energy-diagonal representation.}
\label{tab:natfit}
\begin{center} 
\begin{tabular} {lcc} \hline \hline
&{\em {Quantum}} &  {\em Semi-Classical} \\ \hline
1. & Matrix elements of $x,p$ & Fourier components of $x(I,\theta),p(I,\theta)$ \\
& $ \langle n|x|m\rangle$, $\langle n|p|m\rangle$ & 
$ x_k (I)$, $p_k (I)$ 
%,$x_k (I)=(2\pi)^{-1}\int_{0}^{2\pi}x(I,\theta)e^{-ik\theta} $ 
\\  
2. & Trace variational principle & Constrained variational principle\\
& $\delta {\rm Tr}(H-ip[H,x])=0$  & $\delta \langle (H-\omega p
\frac{\partial x} {\partial \theta})\rangle =0 $ \\ 
3. & Matrix element of EOM & Fourier component of EOM\\  
& $(E_{n+k}-E_n )^2 x_{n+k,n}=\left(\frac{dV}{dx}\right)_{n+k,k} $
& $(k\omega)^2 x_k = \frac{d\langle V\rangle}{dx_{-k}}$ \\   
4. & Commutation relations  &   Quantization of the action \\
& $ \langle n|[x,p]|n\rangle =i$ & $ I=\langle p\frac{\partial x}{\partial 
\theta}\rangle =n + \frac{1}{2} $ \\
5. & Energy is diagonal element of $H$ & Energy is phase average of $H$\\
& $ E_n = \langle n|H|n\rangle $ & $ E(I) =\langle H \rangle $ \\
6. & Quantum frequency    &    Classical frequency   \\
& $ \omega_{n+k,n} \equiv E_{n+k}-E_n $ & $\omega= dE(I)/dI $ \\
7. & Operator          &    Dynamical variable  \\
&   $ A(x,p) $      &  $   A(x(I,\theta),p(I,\theta)) $   \\
8. & Matrix elements  &   Fourier coefficients  \\
& $ \langle n|A|n+k\rangle $  & $ A_k (I) =\int \frac{d\theta }{2\pi}
A(I,\theta)\exp(-ik\theta) $ \\
9.  & Energy is diagonal    & Hamiltonian independent of angles\\
& $ \sum_{k\neq 0}\langle n|H|n+k\rangle\langle n+k|H|n\rangle =0 $ &  
$ \sum_{k\neq 0} H_k H_{-k} =0 $ \\ 
\hline \hline
\end{tabular}
\end{center}
\end{table}

\section{Further study of the commutation relations}  

\subsection{One-dimension and WKB quantization rule}

In the following lines we present a derivation of the WKB quantization rule 
for bound states that is simpler than the one that we have previously published
\cite{15}.  

A diagonal element in a state $n$ of the canonical commutation relation
may be written (using reality conditions)
\begin{equation}
\sum_{n' =0}^{\infty}2x_{nn'}ip_{nn'} =1.  \label{eq:cr1}
\end{equation}
Change $n\rightarrow n''$ and sum this equation from 0 to $n$.  Divide the resulting
double sum into two terms.  The first, defined by taking both sums from
0 to $n$, vanishes by antisymmetry (the matrix elements of $x$ are 
symmetric under exchange of indices, those of $p$ antisymmetric). 
We are left with the sum                 \begin{equation}
\sum_{n''=0}^{n}\sum_{n'=n+1}^{\infty}2x_{n''n'}
ip_{n''n'} =n+1.    \label{eq:cr2}      \end{equation}
With our standard definition,                 \begin{equation}
x_{n'-n}(\frac{1}{2}(n+n'))\equiv x_{nn'},   \label{eq:cr3}
\end{equation}
and a change of indices           \begin{equation}
n''=n-\nu, \;\;n'=n+k-\nu,   \label{eq:cr4}    \end{equation}
(\ref{eq:cr2}) becomes        \begin{equation}
2\sum_{k=1}^{\infty}\sum_{\nu=0}^{k-1}x_k(n+\frac{1}{2}k -\nu)
ip_k(n+\frac{1}{2}k -\nu) = n+1.    \label{eq:cr5}  \end{equation}

Equation (\ref{eq:cr5}) is still exact and can used in quantum calculations.
Expanding about the semiclassical value $x_k(n)$ and keeping terms that 
contribute at most to order $n$ and to order unity, we find 
\begin{equation}
\sum_{k=1}^{\infty}\left[2kx_k(n)ip_k(n) +\frac{d}{dn}
\left(kx_k(n)ip_k(n)\right) +...\right] = n+1.  \label{eq:cr6}
\end{equation}
(To obtain the form of the second term on the left hand side of this equation
requires a careful examination and grouping of terms that contribute.)  
The way to read this equation is to understand that the first term
contains contributions of order $n$ and smaller, whereas the second
term is at most of order unity, and at the same time is half the derivative
with respect to $n$ of the first term.  It follows that this second term  
has the value $\frac{1}{2}$.  We thus derive the one-dimensional Cartesian
WKB quantization rule 
\begin{equation}
2\sum_{k=1}^{\infty}kx_k(n)ip_k(n) = n + \frac{1}{2}.   \label{eq:cr7}
\end{equation}  

A similar tour de force does not work in two or more dimensions.
We omit the uninspiring details.
The moral of the story is that we are "lucky" to have the EBK quantization
rules for semiclassical quantization.  Our aim when we started the investigation
of this section was to discover an alternative semiclassical scheme based
directly on the commutation relations.  The conclusion,
which we believe to be firm, is that
this is not possible except in the one-dimensional case.

\subsection{Remarks on quantum aspects}

Consider a two-dimensional system and suppose it to possess
only bound states.  (In any event   
we use a notation with discrete labels only.)   We can write a Heisenberg
scheme that should be valid whether we deal with the "regular" spectrum
or the "irregular" one \cite{13}.  
We simply order the energy levels of a two-dimensional system with 
coordinates $x,y$, in a linear 
sequence, $N=0,1,...$ .   A calculation can be based on the equations
\begin{eqnarray}
(E_{N'}-E_N)^2 x_{NN'}&=& -(F_x)_{NN'},   \nonumber \\
(E_{N'}-E_N)^2 y_{NN'}&=& -(F_y)_{NN'},   \label{eq:h1} \\  
\sum_{N'}(E_{N'}-E_N)|x_{NN'}|^2 &=& 1,   \nonumber  \\  
\sum_{N'}(E_{N'}-E_N)|y_{NN'}|^2 &=& 1,   \label{eq:h2}  \\  
\end{eqnarray}  
where $E_N$ are the exact eigenvalues.  These equations can be viewed in 
two ways, either as a set of sum rules to be satisfied by the exact
solutions of the quantum mechanical problem, found by some other means
such as matrix diagonalization in a basis, or else as the  foundation 
for a computational scheme involving the solution of non-linear equations.
For a well-defined quantum system (Hamiltonian bounded from below),
we shall certainly be able to do a diagonalization and subsequently check 
the above equations.
Since this should be possible both for the regular and for the irregular
spectrum, this raises the hope that some version of the quantum matrix
method can also be applied to the chaotic regime.  

toward this end, we note that by summing the commutation relations 
from 0 to $N$, we can replace (\ref{eq:h2}) by the positive sums
(this involves the same argument as in the one-dimensional case)
\begin{eqnarray}   
\sum_{N'=N+1}^{\infty}\sum_{N''=0}^{N}(E_{N'}-E_{N''})
|x_{N''N'}|^2 &=& N+1 ,  \nonumber \\
\sum_{N'=N+1}^{\infty}\sum_{N''=0}^{N}(E_{N'}-E_{N''})
|y_{N''N'}|^2 &=& N+1 .  \label{eq:h3}    \end{eqnarray}
These relations are interesting
because they guarantee the convergence of certain sums, and thus 
imply that the matrix elements cannot spread out too far as a function
of energy differences.    

\section{Application of matrix mechanics and comparison with semiclassical approximation}

In this section we present illustrative applications of the matrix
mechanics method for several simple models and compare the results with
those of the semiclassical approximation as well as with the results
of exact diagonalization.  With a given semiclassical approximation,
as defined below,
we shall associate the matrix method, that will be described 
as a sequence of approximations
that should approach more and more closely to the exact result.  Therefore
we may anticipate that at a sufficiently high order of approximation 
its accuracy will exceed that of the semiclassical result.  It is not, however,
guaranteed to do this automatically for the simple reason that the 
quantum method that we shall describe does not have a true semiclassical
limit, but is specifically designed to be most accurate for small quantum
numbers.  For this reason, we shall also describe
an alternative method that starts with the EBK theory and
specifically calculates quantum corrections to it. Numerical results for
all three methods are presented at the end of this section.

Imagine that we have a two-dimensional system described by a Hamiltonian
\begin{eqnarray}
H &=& \frac{p_1^2}{2} + \frac{p_2^2}{2} + \frac{1}{2} \lambda_1 (x^1)^2 +
\frac{1}{2} \lambda_2 (x^2)^2 + V^{\rm anharmonic}(x^1, x^2),
\nonumber \\
{}& \equiv & \frac{{\bf p}^2}{2} + \frac{|{\lambda} \cdot {\bf x}|^2}{2} +
V^{\rm anharmonic}({\bf x}).
\end{eqnarray}
However, the methods described here are more general, and can be used with
a system with many degrees of freedom.  We shall describe these methods, in 
general, by referring to an $n$-dimensional system and give examples
and show results for the specific cases $n=1$ and $n=2$.  For improved
clarity, we have shifted the coordinate index from subscript to superscript.

For the solution of the algebraic formulation of the semiclassical 
approximation we utilize an iterative method 
described by Percival and Pomphrey \cite{11,12}.  For the corresponding 
fully quantum calculations, we shall describe a quantum 
extension of this method.
We have also carried out calculations using the Newton-Raphson method,
but these will not be discussed here.

\subsection{Semi-Classical Iterative Method}

Following Percival and Pomphrey,  
this method utilizes the semi-classical equations of motion, given for Fourier
components ${\bf k}$  (in terms of the above Hamiltonian):
\begin{equation}
\left[ {\lambda} - \left( {\bf k} \cdot {\omega} \right)^2 \right]
{\bf x}_{\bf k} = {\bf F}_{\bf k}(\bf x),
\end{equation}
where
\begin{equation}
{\bf F}({\bf x}) = - \nabla V^{\rm anharmonic}({\bf x}),
\end{equation}
and ${{\omega}=(\omega_1,\omega_2,\ldots,\omega_n)}$ is the vector of
frequencies of the system.  When there is no anharmonic driving force,
$\omega_i^2 = \lambda_i$.
The vector $\bf k$ is an $n$-dimensional sequence
of integers, positive and negative: ${{\bf k}=(k_1,k_2,\ldots,k_n)}$.
It is also useful, here, to define the vectors ${{\bf 1}=(1,0,0,\ldots,0)}$,
${{\bf 2}=(0,1,0,\ldots,0)}$, ${{\bf i} = (0,\ldots,1,\ldots,0)}$, where the
``1'' is in the $i$'th place and all other indices are zero.
Thus, ${\bf k} \cdot {\bf i} = k_i$.  In order to solve the equations that
describe the semiclassical limit in algebraic terms, we must truncate the
Fourier series by specifying a maximum integer vector ${\bf |K|}$.  The meaning
of the absolute value is that we define this limit symmetrically with 
respect to the origin in ${\bf k}$ space.  We set $x^i_{\bf k}=0$
for $|{\bf k}|>|{\bf K}|$.

The procedure begins from the harmonic limit, where the only nonzero Fourier
components are $x^i_{\pm \bf i}$. 
As an example, for a two-dimensional system, where we write $x^1=x$
and $x^2=y$, the 
only nonzero components, initially, are $x_{\pm 1,0}$ and
$y_{0,\pm 1}$.  All others vanish. 
We shall refer to these as the driving components.
The values of these initial nonzero components are computed from the
action conditions, Eq.~(\ref{eq:4.8}),  namely, $x^i_{\pm \bf i} 
= \sqrt{I_i/2\omega_i}$, where $I_i$ is the action of the $i$'th degree of
freedom, as given by Eq.~(\ref{eq:4.7}).

With these initial values, we can then calculate initial values for the 
Fourier components of the non-harmonic part of the force.
Improvements to the frequencies ${\omega_i}$ are then obtained 
from the equations of motion for the driving components,
\begin{equation}
\omega_i^2 = \lambda_i - \frac{F^i_{\bf i}}
{x^i_{\bf i}}.
\label{eq:wnew}
\end{equation}
With these new and improved frequencies, other Fourier components can
be obtained from the remaining equations of motion.  In particular, we compute
\begin{equation}
x^i_{\bf k} = \frac{ F^i_{\bf k}({\bf x})}
{\lambda^i - \left({\bf k} \cdot {\omega} \right)^2} \qquad 
\mbox{\rm for ${\bf k} \ne \bf i$}.
\label{eq:newsmall}
\end{equation}

This completes one cycle of iteration.  We now return to the EBK
quantum conditions, and using the calculated values of the 
frequencies and of the
non-driving Fourier components we compute new and improved values for 
the driving components.  Explicitly
\begin{equation}
x^i_{\bf i} = \sqrt{\frac{1}{2\omega_i} [I_i - 
 \sum_{{\bf k}\neq{\bf i}}k_i({\bf k}\cdot{\bbox \omega)}{\bf x}_{{\bf k}}
{\bf x}_{{\bf -k}}]}.
\end{equation}
For the special case  $n=2$, we have
\begin{eqnarray}
x_{\pm 1, 0} &=& \sqrt{ \frac{I_x - \sum_{(k_1,k_2) \ne (\pm 1,0)} k_1
(x_{k_1,k_2}^2 + y_{k_1,k_2}^2) \left( k_1 \omega_1 +  k_2 \omega_2 \right) }
{2\omega_1}} ,  \\
y_{0,\pm 1} &=& \sqrt{ \frac{I_y - \sum_{(k_1,k_2) \ne (0,\pm 1)} k_2
(x_{k_1,k_2}^2 + y_{k_1,k_2}^2) \left( k_1 \omega_2 +  k_2 \omega_2 \right) }
{2\omega_2}}.
\end{eqnarray}
Once more we calculate the improved values of the frequencies and
then of the remaining components and continue to loop until the energy
converges.

\subsection{Quantum Iterative Method}

In this section, we discuss the modifications to the semiclassical perturbative
method which are necessary in order to have a completely quantum calculation.
The major complication of the quantum approach is that in our version 
of the Heisenberg method, we must consider the coupling of eigenstates
both in the equations of motion and in the commutation relations.  To deal
with this problem in a practical way will necessitate the introduction of
two cut-off parameters, compared to the one used in the semiclassical case.
Consider a typical matrix element,
$\langle {\bf m}|O|{\bf n}\rangle$, of an operator $O$ (in general a 
coordinate).  If we have chosen an ordering of the integer vectors, we can
suppose that ${\bf n}>{\bf m}$ is defined and ${\bf k}\equiv{\bf n}-{\bf m}$
is the analogue of a Fourier component.  Since the first state always
considered in the quantum method to be described 
is the ground state ${\bf m}=0$, 
${\bf K}$ coincides with the maximum value of ${\bf n}$ that we include in the
calculation.  However, this specification does not completely define the 
calculation.  The crudest way to define a closed scheme for fixed 
${\bf K}$ is to allow only matrix elements with ${\bf m}=0$.  
But the equations of motion for the 
matrix elements $\langle {\bf 0}|O|{\bf n}\rangle$ will necessarily
bring in matrix elements with ${\bf m}\neq {\bf 0}$.  In order to have a
closed scheme involving only the quantities specified, we must introduce
a closure approximation that relates the unwanted matrix elements to 
those that belong to the allowed set. The crudest 
such approximation is to relate matrix elements by an equal displacement
of both sets of quantum numbers, i.\ e.,
\begin{equation}
\langle {\bf m}|O|{\bf n}\rangle \cong \langle{\bf 0}|O|{\bf n-m}\rangle.
\label{eq:close} \end{equation}
(A better approximation at this point is to use harmonic oscillator results
for the large matrix elements that are non-vanishing in 
the uncoupled limit.)  This 
crudest of closure approximations is called the zero band width ($B=0$)
approximation.  

To improve the scheme, we increase the band-width $B$.  Thus in a $B=1$
approximation, we treat without closure those values of ${\bf m}$ 
that comprise both the ground state and those states that are coupled to
the ground state by harmonic oscillator matrix elements.  By suitable
extension we can define a scheme for arbitrary $B$, with the understanding
that the maximum value of ${\bf m}$, called ${\bf M}$,  cannot exceed
${\bf K}$.  A closure approximation, generalizing (\ref{eq:close}), needs to
be applied now only to relate matrix elements outside the expanded scheme
to those in which the left hand state is ${\bf M}$.  As $B$ increases, we find
that the values obtained for low-lying states become more and more 
accurate, as well as insensitive to the closure approximation. 
The values of $B$
and of ${\bf K}$ determine the number of equations of motion and
commutation relations that have to be used to evaluate the set of 
included matrix elements and energy differences.   This 
number is a small multiple of $B$ times the number of equations for the 
semiclassical approximation associated with the same value of ${\bf K}$.

Turning to the detailed scheme, we begin from the harmonic limit.  
In this limit, all the energies are
equally spaced.  Thus, if we define quantum frequencies to be the
energy differences
\begin{eqnarray}
\langle {\bf n} | H | {\bf n} \rangle  - \langle {\bf m} | H | {\bf m} \rangle
&=& E_{\bf n} - E_{\bf m}, \nonumber \\
&\equiv& \omega_{ {\bf n}, {\bf m} },
\end{eqnarray}
then the harmonic limit is defined by
\begin{equation}
\omega_{{\bf n} \pm {\bf i}, {\bf n}} = \pm \sqrt{\lambda_i}. \label{eq:freq}
\end{equation}
Here, the vector ${\bf i}$ is the same as was introduced above, namely,
it is an $n$-dimensional vector whose entries are all zero except for the
$i$'th one, whose value is unity.

From the well-known solution for the harmonic oscillator, we also know
the driving matrix elements $x^i_{{\bf n+i,n}}$.  It is integral to our method
that this result need not be put in from the outside but follows by application
of the canonical commutation relations, given the result expressed by 
(\ref{eq:freq}).  Actually the full statement is that for the harmonic case,
as is well known, the simultaneous solution of the equations of motion and of 
the commutation relations yields the starting values of the quantum frequencies
and of the driving matrix elements.  In turn, we can calculate the lowest
approximation to the force $F^i_{{\bf n, n+i}}$.  This completes a single
cycle of the quantum iteration procedure.

Next, the quantum frequencies are improved by using the generalization of
Eq.~(\ref{eq:wnew}):
\begin{equation}
\omega^2_{{\bf n} + {\bf i}, {\bf n}} = \lambda_i 
- \frac{ F^i_{{\bf n}, {\bf n} + {\bf i}}} {x^i_{{\bf n}, {\bf n} + {\bf i}}}.
\end{equation}

For further work, we write 
\begin{eqnarray}
\omega_{{\bf n,m}}&=&\omega_{{\bf m+a,m}}, \\
{\bf a}&=& a_1{\bf 1}+a_2{\bf 2} +...+a_n{\bf n}.
\end{eqnarray}
It is straightforward to write this energy difference as a sum of 
fundamental energy differences or quantum frequencies in which just one of the 
quantum numbers changes by a unit.  This is important in counting that the
number of variables in the problem is determined by the equations of motion
and a set of diagonal elements of the commutation relations.

With the new and improved energy differences, it is possible to compute
the perturbative quantum matrix elements from the equations of motion:
\begin{equation}
x^i_{{\bf n},{\bf n}+{\bf a}} = \frac{F^i_{{\bf n},{\bf n}+{\bf a}}}{
\omega_{{\bf n}+{\bf a},{\bf n}}^2} \qquad
\mbox{\rm for ${\bf a} \ne {\bf i}$}.
\end{equation}

Lastly, we construct new ``fundamental'' matrix elements from the
diagonal matrix elements of the canonical commutation relations.  
Toward this end, we solve the $n$ equations $(i=1,\ldots,n)$
\begin{equation}
\sum_{\bf m} \left( x^i \right)_{{\bf j},{\bf m}}^2
\omega_{{\bf m},{\bf j}} = \frac{1}{2}
\end{equation}
for $x^i_{{\bf m}+{\bf i},{\bf m}}$.
These equations replace the quantum conditions for the action variables,
used in the semi-classical approximation, and, 
as already stated, play the same role as the
latter did, in providing additional equations needed to have a determined
set. 

We have thus identified the three distinct elements that are used
in an iteration cycle:
the commutation relations, the equations of motion for the ``driving"
matrix elements, and the equations of motion for the ``small" matrix elements.

\subsection{Quantum corrections to the semiclassical approximation}

We start by reminding the reader that the semiclassical approximation to the 
energy, $E(n)=\langle n|H|n\rangle$ (in a one-dimensional notation), 
is obtained by expanding energy differences and quantum matrix elements
about certain values that are identified as semiclassical ones and
dropping higher order terms.  With the definitions
\begin{eqnarray}
\omega(n) &=& E(n+\frac{1}{2})-E(n-\frac{1}{2}), \label{eq:omega} \\
k &=& n''-n', \\
\bar{n} &=& \frac{1}{2}(n' + n''),
\end{eqnarray}
we have to second order 
\begin{eqnarray}
\langle n'|x|n''\rangle &=& x_k(n) +(\bar{n}-n)\frac{dx_k(n)}{dn}
+\frac{1}{2}(\bar{n}-n)^2 \frac{d^2 x_k(n)}{d^2 n}, \label{eq:E1} \\
E(n+k)-E(n) &=& k\omega(n) +\frac{1}{2}k^2\frac{d\omega(n)}{dn}
+(\frac{1}{6}k^3 -\frac{1}{24}k)\frac{d^2 \omega(n)}{d^2 n},\label{eq:E2}
\end{eqnarray}
where the derivation of the second of these equations requires attention to 
the definition (\ref{eq:omega}).

To obtain quantum corrections to the energy, we must retain 
and compute the values
of the first and second derivative terms in (\ref{eq:E1}) and (\ref{eq:E2}).
These equations are expansions in $(1/n)$ and we need terms up to second order
because, as we have shown, the first order terms do not contribute to the
energy.

The calculation is straightforward.  Reverting to an $n$-dimensional 
notation, we recall that the semiclassical quantization program is 
based on the equations
\begin{eqnarray}
({\bf k}\cdot\bbox\omega)^2 x^i_k&=& F^i_k, \label{eq:ls1} \\
\sum_{j,{\bf k}}k_i({\bf k}\cdot\bbox\omega)x^j_k x^j_{-k}&=&(n_i 
+ \frac{1}{4}\alpha_i),
\label{eq:ls2}     \end{eqnarray}
where $F_k^i$ is a Fourier component of the $ith$ component of the 
force and $\alpha_i$ is a 
Maslov index.  Let us write these equations succinctly as
\begin{equation}
\Psi_a(z) =0,   \label{eq:ls3}   \end{equation}
where $a$ is a label that enumerates in turn the equation set (\ref{eq:ls1})
and (\ref{eq:ls2}), putting them in one to one correspondence with the Fourier 
components and with the frequencies.  
Correspondingly, $z=\{z_a\}=\{x_k^i,\omega_i\}$.

We now treat these equations as if they are true for a continuous range of 
$n$ and compute the first and second derivatives, leading to the equations,
\begin{eqnarray}
\frac{\partial \Psi_a}{\partial z_b}\frac{\partial z_b}{\partial n_i}
&=& s_{ai},     \label{eq:E3} \\
\frac{\partial \Psi_a}{\partial z_b}\frac{\partial^2 z_b}
{\partial n_i \partial n_j} &=& -\frac{\partial^2 \Psi_a}{\partial z_b
\partial z_c}\frac{\partial z_b}{\partial n_i}\frac{\partial z_c}
{\partial n_j}.  \label{eq:E4}
\end{eqnarray}
Here the column matrix $s_{ai}$, consisting of all zeros
and a single value of unity has its origin in the right hand side of 
Eq.\ (\ref{eq:ls2}). 
Both of the above sets of equations are linear and inhomogeneous and will have 
a solution provided the matrix
\begin{equation}
S_{a,b} = \frac{\partial \Psi_a}{\partial z_b}  \label{eq:E5}
\end{equation}
is non-singular.  (See the next section for further discussion of the 
uses of this matrix.)  
The solutions of these equations provides us with the
necessary input for calculating the leading quantum corrections to the 
energy.

\subsection{Numerical illustrations}

We consider first the one-dimensional inverted quartic potential.
\begin{equation}
H= \frac{1}{2}(p^2 + x^2) +bx^4,  \label{eq:EX1}
\end{equation}
with $b<0$.  For this model as well as for the two dimensional model 
considered below, it is understood that these do not exist as quantum mechanical systems.  Nevertheless diagonalization in a large - but not too large -
basis will yield eigenvalues that we can treat as those of a bound system.
Correspondingly the approximation methods described above will also yield
values that we can take seriously and compare with the ``exact" results.
In the following and for the two dimensional model, we shall present a
few examples of the calculations that we have done.

In Table II we display the energies of the ground state and of two excited
states calculated for several values of $b$, both for the inverted 
quartic and the stable quartic oscillator.  Four different calculations
have been carried out at each point:  $E_{\rm diag}$ results from matrix
diagonalization (``exact'' value); $E_{\rm sc}$ is the semiclassical
result; $E_{\rm sc}+\Delta E$ is the semiclassical result with leading
quantum corrections; $E_q$ is the value obtained from Heisenberg matrix
mechanics.  For the latter, the band width was chosen, when feasible, to give 
results in essential agreement with $E_{\rm diag}$.  This required values of
$B=3-5$.
Our general expectation is that the semiclassical
calculation with quantum corrections will be an improvement over the purely
semiclassical calculation.  This expectation is born out 
with the exception of the value for $b=-.005$, $n=10$.  
Since the maximum of the potential for $b < 0$  is at ~$\frac{1}{16 |b|}
=12.5$ for $b=-0.005$, the $n=10$ level gets sufficiently close
to this maximum that the assumptions upon which our derivation of the 
semiclassical approximation and of the quantum corrections to it are based
become suspect.  It was to illustrate this point that results for
the stable quartic potential were included.  Here there is no sign of the 
difficulty encountered with the inverted potential.

\begin{table}
\caption[]{Representative values of the energy for states $n$ of 
a one dimensional quartic oscillator, for several values of the 
anharmonicity parameter $b$.
 $E_{i\rm diag}$ denotes the energy
calculated by direct diagonalization, $E_{\rm sc}$ is the semi-classical
approximation, $E_{\rm sc}+ \Delta E$ includes the leading quantum correction
to the semiclassical result,
and $E_{q}$ is the quantum result obtained by the Heisenberg method.}
\begin{tabular}{dddddd} \hline 
{\it b} & $n$ & $E_{\rm diag}$ & $E_{\rm sc}$ & $E_{\rm sc}+ \Delta E$ 
& $E_{q}$ 
\\ \hline
      & 0 & 0.498489 & 0.499248 & 0.498500 & 0.498489 
\\
-0.002  & 5 & 5.405378 & 5.406235 & 5.405521 & 5.405377 
\\
      & 10 & 10.145887 & 10.146877 & 10.146244 & 10.145887 
\\ 
\hline
      & 0 & 0.496182 & 0.498112 & 0.496249 & 0.496182 
\\
-0.005  & 5 & 5.249313 & 5.252083 & 5.250752 & 5.249314 
\\ 
      & 10 & 9.484437 & 9.489854  & 9.493750 & 9.484438 
\\
\hline
      & 0 & 0.501490 & 0.500748 & 0.501500 & 0.501490 
\\
+0.002  & 5 & 5.588750 & 5.588081 & 5.588840 & 5.588750 
\\
      & 10 & 10.813669 & 10.813058 & 10.813814 & 10.813668 
\\ 
\hline
      & 0 & 0.503687 & 0.501862 & 0.503746 & 0.503687 
\\
+0.005  & 5 & 5.712896 & 5.711448 & 5.713323 & 5.712896 
\\ 
      & 10 & 11.231303 & 11.230091  & 11.231885 & 11.231304 
\\
\hline
\end{tabular}
\end{table}

In Fig.\ 1 we compare, as a function of the coupling constant,
semiclassical matrix elements with the two
neighboring quantum matrix elements of which it should be the 
approximate average. (The notation makes it clear that these results are 
for the reference state $n=10$.)  The expectation is well satisfied for 
the largest matrix elements.  We also see that except for the fundamental
Fourier component, all other matrix elements are rising exponentially
with increasing coupling constant, promising an eventual breakdown of the 
theory.

\begin{figure}[]
	\centerline{\psfig{figure=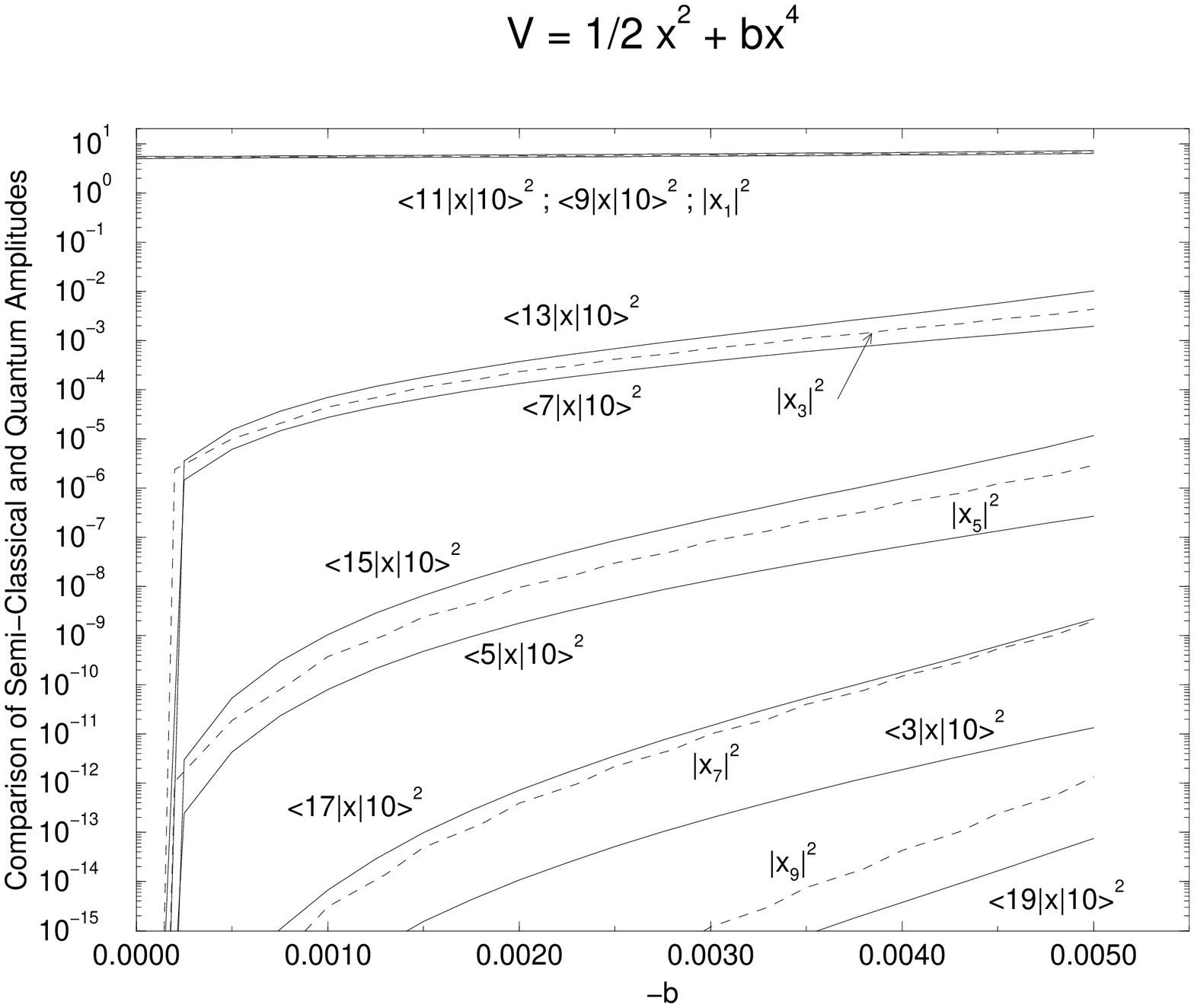,height=4.0in,angle=-90}}
	\caption[]{Comparison of semiclassical matrix elements 
(Fourier components) for the inverted quartic oscillator and the 
reference state $n=10$ with the quantum matrix elements that they most 
closely approximate, as a function of the anharmonicity parameter $b$.
The square of the value is plotted with the semiclassical result shown as
as a dashed line and the associated quantum values as full lines.}
	\label{}
\end{figure}

We turn next to the two dimensional generalized Henon-Heiles model,
\begin{eqnarray}
H&=&\frac{1}{2}(p_x^2 +\lambda x^2) +\frac{1}{2}(p_y^2 + \mu y^2)
+V(x,y), \\
V(x,y) &=& by(x^2 +b'y^2),
\end{eqnarray}
where we choose $\lambda=1.69$ and $\mu=.49$, in order to stay away from
low-order resonances. We shall consider a range of values of $b$
and $b'$ in order to expose different physical situations.

In Table III we display a limited number of results analogous to those
presented in Table II.  Here the failure of the quantum corrections to
the semiclassical result is flagrant for the first choice of interaction
parameters, which can be ascribed to the smallness of the height of the 
lowest barrier.  This barrier is raised by the second choice of parameters,
and this seems to cure the difficulty.  Again, the accuracy of the non-linear
calculations carried out for advertised form of Heisenberg matrix mechanics
is to be noted.

\begin{table}
\caption[]{ Energies of states labeled by  $n_1,n_2$ 
for the Henon-Heiles model defined in the text, as a function of the 
anharmonicity parameter $b$, for a fixed value of $b'$. 
$E_{\rm diag}$ denotes the energy
calculated by direct diagonalization, $E_{\rm sc}$ is the semi-classical
approximation, $E_{\rm sc}+ \Delta E$ the quantum-corrected semiclassical
approximation, and $E_{q}$ is the quantum result obtained by the Heisenberg
method.}

\begin{tabular}{dddddd} \hline 
{\it b ~ b'} & $n_1~n_2$ & $E_{diag}$ & $E_{sc}$ & $E_{sc}+ \Delta E$ & $E_{q}$ 
\\ \hline
      	    & 0~ 0 & 0.999628 & 0.999627 & 0.999757 & 0.999628 
\\
-0.04~ 0.01 & 2~ 2 & 4.989952 & 4.990620 & 4.991267 & 4.989952 
\\
            & 4~ 4 & 8.967166 & 8.969413  & 8.969985 & 8.967164
\\ 
\hline
             & 0~ 0 & 0.998504 & 0.998501 & 0.998817 & 0.998504  
\\
-0.08~ 0.01  & 2~ 2 & 4.958779 & 4.961547 & 4.961127 & 4.958780 
\\ 
             & 4~ 4 & 8.862238 & 8.871849 & 8.862532 & 8.862098 
\\
\hline
\end{tabular}
\end{table}

In Fig.\ 2 and Fig.\ 3 we compare semiclassical and quantum matrix elements
for the coordinates $x$ and $y$, respectively, in analogy with what is shown
in the one dimensional case.  The same general remarks apply here as 
for that case.  Note that the range of parameters does not include those 
used in Table III.

\begin{figure}[]
	\centerline{\psfig{figure=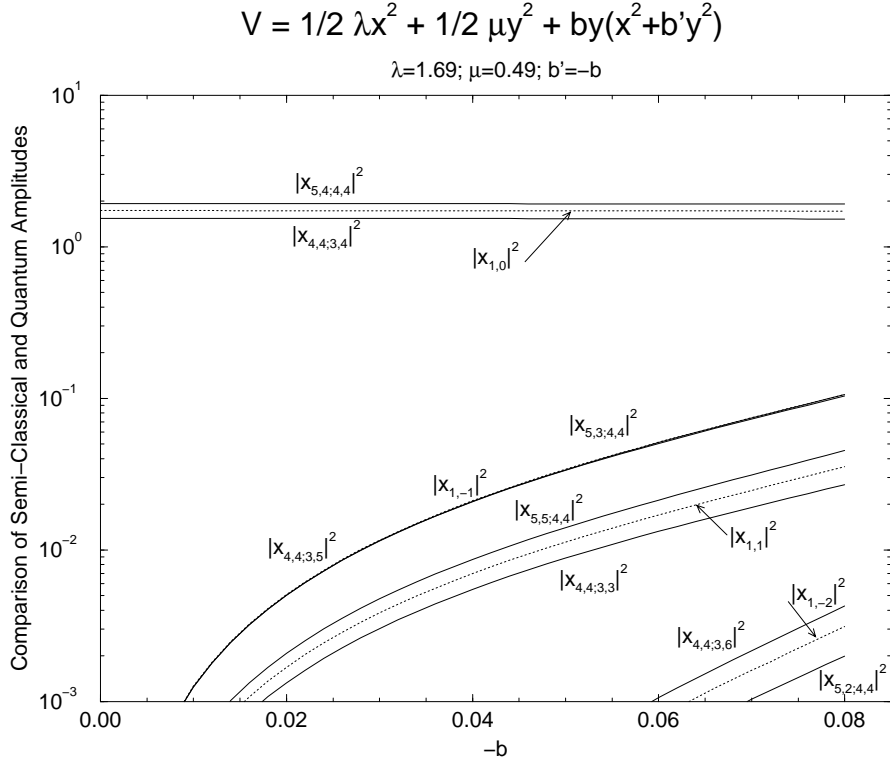,height=4.0in,angle=-90}}
	\caption[]{For the potential shown at the top of the figure,
comparison for the state $(n_1=n_2=4)$ of quantum values
and of the associated semiclassical approximation for 
selected matrix elements of the coordinate $x$. The full lines represent 
the quantum results and the dashed ones the semiclassical values. }
	\label{}
\end{figure}

\begin{figure}[]
	\centerline{\psfig{figure=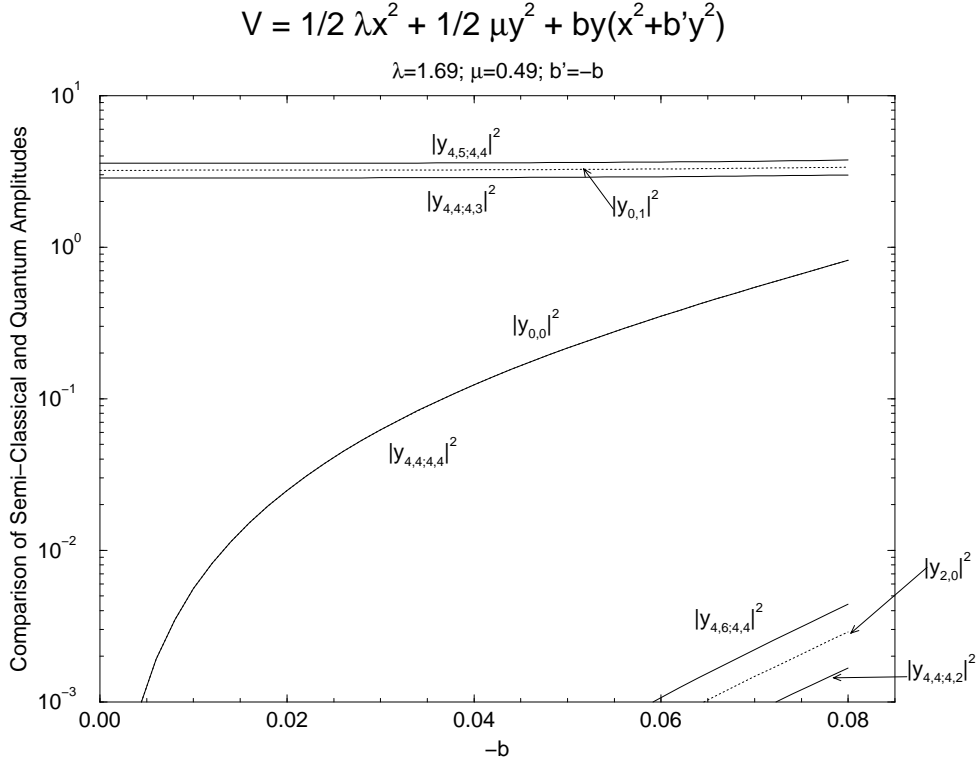,height=4.0in,angle=-90}}
	\caption[]{The same comparison as in Fig.\ 2, for the coordinate 
$y$.}
	\label{}
\end{figure}

\section{Additional observations and suggestions concerning the semiclassical 
and classical analyses}
In this section, we raise a number of issues related to the semiclassical
analysis that it may to profitable to explore.  Since some of the 
suggestions put forward in this section are speculative, they may
also turn out to be worthless.

\subsection{Fixed action quantization program}

We first consider other possible applications of linear variation about
the semiclassical equations, as embodied in Eq.\ (\ref{eq:E3}).  These
equations are a special case of the equations
\begin{equation}
S_{a,b}\delta z_b= s_a, \label{eq:ls4}
\end{equation}
which has a number of possible applications, 
depending on the choice of the source vector
$s$.  We discuss briefly several such applications. \\  

A1)   There is first of all the basic algorithm of the Newton-Raphson method.
In this instance,  
we choose $s_a =-\Psi_a^{(\nu)}$, corresponding to a $\nu th$ approximation
to $z_a$, and the matrix $S$ is also known in this approximation.  The solution
to (\ref{eq:ls4}) determines the next approximation to $z_a$, namely
\begin{equation}
z_a^{(\nu +1)} =z_a^{(\nu)}+\delta z_a.    \label{eq:ls5}  \end{equation}  

A2) If the above solution converges, we thereby define a "stability 
matrix", $S$.  As we have already seen, it is this quantity that 
occurs when we differentiate
Eqs.\ (\ref{eq:ls1}) and (\ref{eq:ls2}) with respect to $n_i$, leading to 
Eqs.\ (\ref{eq:E3}) and (\ref{eq:E4}).  
In addition to the application already described in the previous section, we 
can think of at several other applications of these equations.  We consider
only (\ref{eq:E3}), but (\ref{eq:E4}) could be brought in.
For example, they allow us to compute the energy of neighboring states,
according to the formula (two-dimensional example),
\begin{equation}
E(n_1+d_1,n_2+d_2) =E(n_1,n_2) +\omega_id_i +\frac{1}{2}\omega_{ij}d_id_j,
\label{eq:ls6}      \end{equation}
where $\omega_{ij}=(\partial\omega_i/\partial n_j)$.
 Work found in the literature, as far as we are aware, uses only the
linear approximation to this result.
As a second application, when we want to calculate solutions for a fixed 
Hamiltonian and for values of the action neighboring to those for which
solutions are already known, the solutions of the linearized equations
can be used to obtain improved starting values for the Newton-Raphson
iteration.\\

A3) To obtain solutions for a neighboring Hamiltonian, one can again use
a form of (\ref{eq:ls4}), with yet another driving term easily derived from
the structure of the original nonlinear equations.

An important question that should be susceptible to study by the 
linearized formalism is the relation of an instability of a solution
of the nonlinear equations to the eigenvalues of the matrix $S$.  

\subsection{Solutions at fixed frequency; applications}  

We now restrict our attention to the solutions of the equations 
of motion (\ref{eq:ls1})
for fixed frequency, i.\ e., we study the purely classical problem,
setting aside for the moment the question of how to adjoin a quantization
feature.  Let us imagine that we are interested in obtaining the Fourier
coefficients as functions of the frequency values. 
There arises the practical problem of how to choose a sensible
grid of frequency values.  This problem can be solved presumably by   
perturbation theory for small coupling, and we can use a form of the   
linearized analysis described above to show 
us how to change frequencies locally.
As a first approximation to quantization - if only energies are of interest -
we can compute values of the action from the Fourier coefficients and the
frequencies and write            \begin{equation}
\tilde{J}_i = r_i +\frac{1}{2} ,    \label{eq:ls7}       \end{equation}
where we have used the tilde to indicate a non-quantized value, so that
$r_i$ is not an integer.  We can then extrapolate to the nearest quantized
values of the energy by using Eq.\ (\ref{eq:ls6}).  If we want also
the Fourier amplitudes for the quantized invariant tori, 
we have to successively modify the 
frequencies until they assume values associated with quantized actions.
for this purpose we might use the formula,    \begin{equation}  
\delta\omega_i =\frac{\partial \omega_i}{\partial J_j}\delta J_j    
= \frac{\partial^2 H}{\partial J_i\partial J_j}(n_j - r_j).
\label{eq:ls8}  \end{equation}
The most practical way of calculating the required second derivatives would
probably be from a grid of energy values.  

We finally note another possibly interesting way of utilizing a grid of 
fixed frequency solutions to quantize a system.  Consider, as a simple 
example, a two 
dimensional non-resonant system.  Write the classical Hamiltonian in the 
normal form              \begin{equation}     
H = \sum a_{m_1m_2} J_1^{m_1}J_2^{m_2}.    \label{eq:ls9}     \end{equation}
It would appear at first sight that   
the number of points on our grid of solution values will determine
the number of terms that we can use in this equation, which yields a 
set of linear inhomogeneous values for the coefficients $a_{m_1m_2}$.
We would then quantize the resulting Hamiltonian by using the EBK quantization
conditions or some variant, following the discussion of Ref.\cite{54}. 
If we consider the usual procedure for constructing a form such as 
(\ref{eq:ls9}), however, we would guess that our previous remarks are much
too naive.  One's ability to carry the expansion to a higher and higher
order is contingent upon obtaining a perturbation expansion to the appropriate
order.  To make contact with methods based on Fourier series, we must note
that an analysis can be carried out which informs us which Fourier components
must be included to guarantee equivalence to a perturbation expansion up to 
a prescribed order.  The solution of a nonlinear scheme including only these
components, at the same time that it contains a selective (and uncontrolled)
summation of higher order terms, is at least perturbatively correct to some
controlled order.  It is this latter order which would determine how many
terms are allowed in the expansion (\ref{eq:ls9}).  To our knowledge this
relationship between the methods based on Fourier expansion and those based
on normal forms has not been considered previously.

\subsection{Direct use of the variational principle}

Can one use the variational principle directly to simply solution of the 
nonlinear equations?  Let $\langle L\rangle^{(\nu)}$ be the $\nu th$
approximation to the average Lagrangian.  Expanding to first order,
\begin{equation}
\langle L\rangle =\langle L\rangle^{(\nu)}+ \frac{\delta\langle L
\rangle^{(\nu)}}{\delta x_{i,k}}\delta x_{i,k} +\frac{\delta\langle L 
\rangle^{(\nu)}}{\delta\omega_i}\delta\omega_i . \label{eq:ls10} 
\end{equation}
Consider first the fixed frequency case and choose    \begin{equation}
\delta x_{i,k} =c_\nu \frac{\delta \langle L\rangle^{(\nu)}}{\delta
x_{i,k}}.     \label{eq:ls11}      \end{equation}
This is useful provided the average Lagrangian is truly an extremum. It
would then appear that a suitable choice of the constant $c_\nu$ both as 
to magnitude and sign would move the average Lagrangian to its extreme value.
The reason that we must allow the constant to depend on the order of 
approximation is that we want to guarantee that the retained first order
correction is larger than the omitted second-order terms.  This means
that we must start out with conservatively small values of the constant
and let it increase toward unity as (if) we approach convergence.

If we carry out a calculation at fixed action, we have to consider the 
equation,  \begin{equation}
\frac{\delta\langle L\rangle^{(\nu)}}{\delta \omega_i}=J_i^{(\nu)}
=J_i.      \label{eq:ls12}       \end{equation}
Comparison with (\ref{eq:ls10}) suggests that we choose       \begin{equation}
\delta\omega_j =b_j J_j \;\;(\rm no\; sum).    \label{eq:ls13}  \end{equation}
From this assumption, the assumption of fixed $J$, and (\ref{eq:ls11}) 
we can derive a pair of linear 
equations for $b_j$, namely          \begin{equation}
\sum_j \frac{\partial J_i}{\partial\omega_j}J_j b_j =- \frac{\partial J_i}
{\partial x_{j,k}}\frac{\delta\langle L\rangle}{\delta x_{j,k}} c.
\label{eq:ls14}  \end{equation}
Here indices indicating the order of approximation are omitted.
In the present case, there is no guarantee that the two first order correction
terms in (\ref{eq:ls10})   
are of the same sign, and thus the way that convergence may be achieved,
if at all, is somewhat more problematical.

\section{Concluding remarks}   
                                                                                
In this paper we have presented a new view of the transition from
Heisenberg matrix mechanics to the theory of invariant tori.
We have suggested a number of possible applications of the ideas 
that were presented mainly in Secs.\ II, IV, and VII. 
We have applied the ideas developed in Secs.\ II and IV to several standard
models.  The most important feature that has emerged 
from these applications is that  
the Heisenberg mechanics can be developed into a quantum calculus
that is only modestly more complicated than the semiclassical 
calculations associated with the theory of invariant tori.   
This can be done in two ways, either by starting with the semiclassical
approximation and building a correction scheme about it, 
or else by constructing a fully quantum scheme starting from the ground
state.  

If one were now to ask for the most important next step that 
one could take with the Heisenberg methods, an excellent candidate for an
answer would be to produce solutions for a globally chaotic system such as the 
one studied by Martens {\it et al} \cite{102}.  
It would also be worthwhile to revisit some old ground.  As an example,
we might restudy the edges of the regions, as a function of    
coupling strength and quantized actions,     
beyond which converged solutions of our equations cannot be found.
Though some of this was done in the present work, the issue is somewhat
muddied for the model chosen, since it does not, strictly speaking, 
possess a Hilbert space.   
There is evidence, based on calculations for the 
standard mapping \cite{21,22}, that this failure represents an independent 
method for studying the disappearance  of invariant tori.  

Another idea that might be reexamined is that of approximate tori.  
Reinhardt and associates have championed this idea, that permits    
EBK quantization to be applied past the point of breakdown of the 
associated invariant torus.  This idea is suggested both by their work 
on the quantization of normal-form Hamiltonians \cite{52,53} and on
adiabatic switching \cite{18}.  It reappears in the work of Martens and
Ezra \cite{20,20a} in that trajectories that appear to be associated with
locally chaotic regions show enough well-defined Fourier components
so that quantized actions can be calculated.  Independently, we realized
that this phenomena would manifest itself in our work as follows:  
The non-linear equations underlying our approach yield   
a solution when only a small number of Fourier components are retained, but
the solution blows up when an attempt to add components is made.
By contrast, when invariant tori exist, this is signaled by insensitivity
of the solution to the addition of Fourier components beyond a fixed
number.  The finite Fourier sums describe approximate invariant
tori in the same general sense, though in a different approximation,
as the approximate normal-form Hamiltonians. 

The transition from the quantum to the classical domain by the methods 
of this paper presents a new aspect of the study of the consequences of the 
order in which the two limits $\hbar\rightarrow 0$,
$t\rightarrow \infty$ are taken \cite{26}.  
For the order studied in this paper, 
in which the time limit is taken first, it is quite impossible to strictly
reach the regime of multiply periodic motion, as has already 
been pointed out in the body of our work.  It may be of interest to 
to undertake a further study of the equations that can be obtained in this 
limit.   There may also be 
some connection of these ideas with the idea of approximate tori.

\acknowledgements     
We are grateful to J.-M.\ Yuan and M.\ Valli\`eres for helpful discussions.
This work was supported in part by the United States Department of Energy,
under grant number 40264-5-25351 and by the R.\ O.\ C.\ National Science
Council, through contract NSC83-0208-M-002-016.

\appendix
                                                                                
\section{Proof that solving the equations of motion diagonalizes the 
Hamiltonian}  

We prove directly that a solution of the equations of motion and
of the commutation relations guarantees       
that the Hamiltonian has been rendered diagonal.  From the equations
of motion, we can derive, in a way which is obvious from the right
hand side of what follows, the equation   
\begin{equation}     
\sum_{l} \{ (E_{l} - E_{m})^{2} + (E_{n} - E_{l})^{2} \} x_{ml} x_{lm}          = (xV' + V'x)_{mn}  = 2(xV')_{mn}  . \label{eq:A4}   \end{equation} 
>From (\ref{eq:2.5}) we may write  \begin{equation}                               (p^{2})_{mn} = - \sum_{l} (E_{l} - E_{m}) (E_{n} - E_{l}) x_{ml}x_{ln} .     
\label{eq:A.7}     \end{equation}                                               Equations (\ref{eq:A4}) and (\ref{eq:2.5})                    
can be combined in several useful ways:               
Thus by subtracting twice (\ref{eq:2.5}) from (\ref{eq:A4}) we find              \begin{equation}                  
E_{n} - E_{m})^{2}(x^{2})_{mn} = - 2(p^{2})_{mn} + 2(V'x)_{mn}  , 
\label{eq:A8}   \end{equation} 
which is the matrix element of the equation                                    
\begin{equation}
 \frac{1}{2} [[x^{2},H],H]= - p^{2} + V'x)  . \label{eq:A9} 
\end{equation}

On the other hand by adding twice (\ref{eq:2.5}) and (\ref{eq:A4}), 
we find the results      \begin{equation}     
 \sum_{l}(2E_{l} - E_{m} - E_{n})^{2} x_{ml} x_{ln} = 2(p^{2})_{mn} +         
2(V'x)_{mn}  . \label{eq:A10}     \end{equation} 
Equation (\ref{eq:A10}) will be used to prove that 
H is diagonal in conjunction with     
another relation which is a further consequence of the equations
of motion, namely,
\widetext  
\begin{equation}
(E_{n} - E_{m}) \sum_{l} (2E_{l} - E_{m} - E_{n})^{2} x_{ml}x_{ln}         
 = - 2i(V'p + pV')_{mn} + 2(E_{n} - E_{m}) (V'x)_{mn}.      
\label{eq:A11}       \end{equation}
\narrowtext  
Since    \begin{equation}             
 i(V'p + pV')_{mn} = 2[V,H]_{mn} = 2(E_{n} - E_{m})V_{mn}, \label{eq:A12}
\end{equation}
we have finally, for $m \neq n$, in place of (\ref{eq:A10}),
\begin{equation}
 \sum_{l} (2E_{l} - E_{m} - E_{n})^{2} x_{ml}x_{lm} = -4V_{mn}                
+ 2(V'x)_{mn}. \label{eq:A13}    \end{equation} 

Comparing (\ref{eq:A9}) with (\ref{eq:A13}), we conclude that
\begin{equation}
2(p^2)_{mn}+2(V'x)_{mn}=-4V_{mn} +2(V'x)_{mn}, \label{eq:A14}
\end{equation}   or  
\begin{equation}
 H_{mn} = \frac{1}{2}(p^{2})_{mn} + V_{mn} = 0,  \;\;\; m \neq n  .            
\label{eq:A15}   \end{equation}

\end{document}